\documentclass[12pt,a4paper]{article}
\usepackage{xcolor}
\usepackage{graphicx}
\usepackage{float}
\usepackage[margin=2cm]{geometry}
\usepackage{amsmath}
\usepackage{tikz}
\usepackage{booktabs}
\usepackage{amssymb}
\usepackage[font={small,it}]{caption}
\usepackage{mathtools}
\usepackage{braket}
\usepackage{subcaption}
\usepackage{subfloat}
\usepackage{cleveref}
\usepackage{verbatim}
\usepackage{cite}
\usepackage[labelfont=bf]{caption}
\usepackage{upgreek}
\usepackage{authblk}

\usepackage[separate-uncertainty = true,multi-part-units=single]{siunitx}
\title{Two-photon 3D printing of functional microstructures inside living cells}

\author[1]{Maru\v sa Mur}
\author[1,2]{Alja\v z Kav\v ci\v c}
\author[1]{Uro\v s Jagodi\v c}
\author[1]{Rok Podlipec}
\author[1,2,3]{Matja\v z Humar*}
\affil[1]{Department of Condensed Matter Physics, J. Stefan Institute, Jamova 39, SI-1000 Ljubljana, Slovenia}
\affil[2]{Faculty of Mathematics and Physics, University of Ljubljana, Jadranska 19, SI-1000 Ljubljana, Slovenia}
\affil[3]{CENN Nanocenter, Jamova 39, SI-1000 Ljubljana, Slovenia}

\affil[ ]{*E-mail: matjaz.humar@ijs.si}

\date{}

\begin{document}
\maketitle

\begin{abstract}


3D printing has revolutionized numerous scientific fields and industries \cite{espera20193d, wallin20183d, gonzalez2023micro, wang2023two}, with printing in biological systems emerging as a rapidly advancing area of research \cite{hippler20193d, guvendiren20193d}. However, its application to the subcellular level remains largely unexplored. Here, we demonstrate for the first time the fabrication of custom-shaped polymeric microstructures directly inside living cells using two-photon polymerization \cite{maruo1997three}. A biocompatible photoresist is injected into live cells and selectively polymerized with a femtosecond laser. The unpolymerized photoresist is dissolved naturally within the cytoplasm, leaving behind stable intracellular structures with submicron resolution within live cells. We printed various shapes, including a \SI{10}{\micro m} elephant, barcodes for cell tracking, diffraction gratings for remote readout, and microlasers. Our top-down intracellular biofabrication approach, combined with existing functional photoresists, could open new avenues for various applications, including intracellular sensing, biomechanical manipulation, bioelectronics, and targeted intracellular drug delivery. Moreover, these embedded structures could offer unprecedented control over the intracellular environment, enabling the engineering of cellular properties beyond those found in nature.

\end{abstract}

\newpage
 

Over the past decade, 3D printing has become an indispensable tool in industry and various scientific fields, such as electronics \cite{espera20193d}, soft robotics \cite{wallin20183d}, micro-optics and photonics \cite{gonzalez2023micro,wang2023two}, biology  \cite{hippler20193d}, and biomedicine  \cite{guvendiren20193d}. For nano- and micro-scale structures, various 3D printing techniques can be used. The best printing resolution can be achieved with the two-photon polymerization (TPP) technique \cite{maruo1997three}, where a photo-sensitive resin (a photoresist) is illuminated with a femtosecond laser. Two-photon absorption occurs only in the small volume of the laser-beam focus, where the laser intensity is sufficiently high. This results in the photo-polymerization occurring only in a limited voxel, enabling printing features that are down to \SI{100}{nm} in size.

In biological environments, TPP is performed with bio-compatible photoresists, mostly to print scaffolds onto which cells can later be seeded to achieve tissue growth and assist with tissue regeneration \cite{liao2020material, hippler20193d}. Additionally, cells or bacteria can be added to the photoresist (photo-responsive hydrogel) to print cellular structures of custom shapes \cite{huang2021highly, hasselmann2018attachment}. Lately, 3D printing with TPP has also been shown inside living organisms. 
In one case, a photo-responsive hydrogel ink with or without cells was injected into several tissues inside live mice. Structures \SI{\sim 1}{mm} in size were polymerized through TPP \cite{urciuolo2020intravital}. 
In another example, a droplet of a commercial elastomeric ink was injected into tissue inside a live embryo of a fruit fly \textit{Drosophila melanogaster} or of a medaka fish \textit{Oryzias latipes} \cite{afting2024minimal}. Inside the photoresist droplet, a structure approximately \SIrange{50}{60}{\micro m} in diameter was printed by TPP.  

There has been one report on TPP inside a synthetic cell \cite{abele2022two}, however, we have not found any documented attempt at 3D printing objects inside living cells. Instead, to be employed for different intracellular applications, micro-scale objects usually get embedded into the cells through phagocytosis. Inside cells, they can serve as probes for intracellular microrheology and force measurement with optical  \cite{arbore2019probing} or magnetic tweezers \cite{timonen2017tweezing}, and as sensors for various parameters inside the cell (such as refractive index, pH, temperature, etc.)  \cite{schubert2020monitoring,kavcic2022deep, gomez2013silicon}. Another application is barcoding, where a foreign 3D object is used as an identification tag for a cell it resides in.  By tagging them, the behavior of individual cells can be studied instead of the usual average responses obtained from large cell populations. Tagging and tracking of cells has been reported with micro-particles of different shapes acting as either graphical \cite{fernandez2009intracellular} or spectral barcodes (microlasers) \cite{humar2015intracellular,martino2019wavelength,anwar2023microcavity}.

Here, we demonstrate 3D printing directly inside living cells, laying the groundwork for a new class of intracellular bioengineering tools and applications.

\section*{Printing in living cells}

\begin{figure}[t!]
	\centering
	\includegraphics[width=17cm]{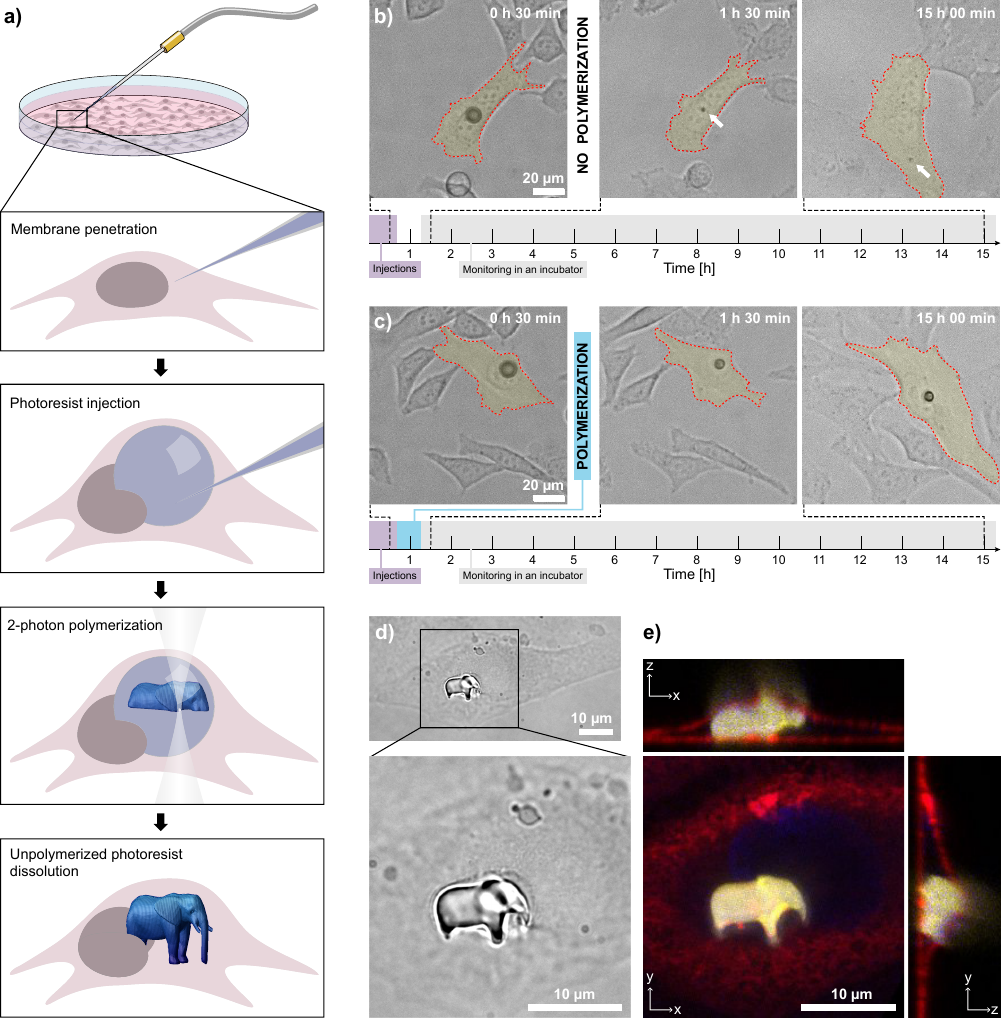}
	\caption{Procedure of 3D printing in cells. (a) Schematic representation of the printing protocol. The photoresist is injected into a cell and illuminated in a pre-designed pattern to form a polymerized structure. (b) A droplet of photoresist, injected into a live HeLa cell, dissolves almost completely in the first two hours after the injection. White arrows point to the photoresist residue \SI{\sim 2}{\micro m} in size, which can sometimes be observed in cells even after a long time. (c) A droplet of photoresist, injected into a live HeLa cell, is completely exposed to laser light \SI{\sim 30}{min} after the injection. After exposure, the photoresist polymerizes and the droplet stops dissolving. (d) Bright-field image of a \SI{10}{\micro m} elephant, printed inside a live HeLa cell. (e) Confocal image of the structure in (d). Cross-sections $xz$ and $yz$ clearly show, that the structure is embedded in the cell, as the membrane (red) is seen covering the elephant (yellow). To improve the structure shape recognition and nucleus visibility in the $xy$-panel, instead of the usual confocal cross-sections the yellow and blue channels show maximum-intensity projections along the z-axis. 
}
	\label{fig:1}
\end{figure}

To print inside cells, a droplet of a negative-tone photoresist was injected into a live HeLa cell (Supplementary Fig.\ 1). The droplet was illuminated in a designed pattern using a commercial system for TPP containing a femtosecond \SI{780}{nm} laser. The cell as well as the medium are transparent for the near-IR laser light. Only in the laser focal spot was the light intensity high enough for the material to be polymerized through the process of TPP. The focal spot is moved layer by layer along the designed path, forming a solid structure of a desired 3D shape inside the cell. The unpolymerized photoresist slowly dissolved, leaving only the polymerized structure in the cell (Fig.\ \ref{fig:1}a). A more detailed experimental timeline is given in Supplementary Fig.\ 2.

The first step of the study was to find a photoresist that is not toxic to the cells. Several commercially available photoresists are reported to be biocompatible when polymerized. However, in our case, the requirements for a photoresist were more strict, since it had to be bio-compatible also in the unpolymerized state. Additionally, it had to be slightly soluble in water, so that the unpolymerized part can be dissolved after some time without additional chemicals. We have searched through several commercial photoresists to identify those that do not appear to be damaging to the cells. We checked IP-S, IP-Visio, IP-L, IP-Dip, and IP-n162 (all by Nanoscribe GmbH). We chose the photoresist IP-S as it impacted the cells the least, while also being slightly soluble in an aqueous medium. A droplet approximately \SI{10}{\micro m} in diameter dissolves entirely (except for a sometimes observed small photoresist residue) in a few hours in an aqueous environment (Fig.\ \ref{fig:1}b). A slight disadvantage of the photoresist solubility is the time limit it sets for the printing: the structures need to be printed within \SIrange{1}{2}{h} after the photoresist droplet is injected, depending on the initial droplet size. Figure \ref{fig:1}c shows an example where a whole droplet was polymerized. After the injection, the droplet is decreasing in size until a certain time-point at which the laser illumination takes place (between the left and middle panels). After that, the droplet diameter remains fixed even after a long time.

In droplets \SIrange{10}{15}{\micro m} in diameter, embedded in cells, we printed various 3D structures approximately \SI{10}{\micro m} in size (Supplementary Fig.\ 3). For example, Fig.\ \ref{fig:1}d shows a bright-field image and its zoom-in of a tiny \SI{10}{\micro m} elephant printed inside a living cell. Additionally, 3D confocal imaging was performed to confirm that the structure does indeed lie inside the cell (Fig.\ \ref{fig:1}e).

\section*{Viability of cells with printed structures}

\begin{figure}[t!]
	\centering
	\includegraphics[width=17cm]{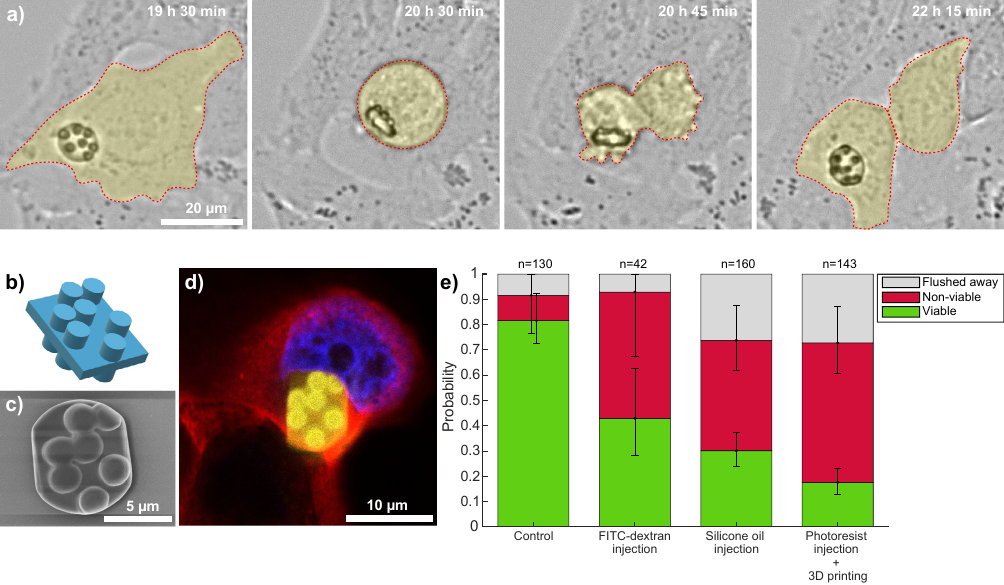}
	\caption{Viability of cells with printed structures. (a) A HeLa cell containing a printed structure undergoes cell division, starting in the second panel, 20.5 hours after injection. (b) A CAD design of the structure - J. Stefan Institute logo. (c) SEM image of the same structure when printed in a droplet of IP-S photoresist. The structure is truncated at the edges since it was slightly larger than the droplet. (d) Confocal image of a HeLa cell, containing the polymerized structure. The cell nucleus visibly deforms and gives way to the structure. (e) In the viability study, HeLa cells were monitored over the course of 24 hours. We show results for cells containing polymerized IP-S structures ($n=143$) and three controls. In the first control, we show cells ($n=130$) that were not manipulated in any way. In the second control ($n=42$), we tested how penetrating the membrane affects the cell: a small amount of FITC-dextran-dyed medium was injected. In the third control ($n=160$), we tested how a sudden increase in the cell volume due to a droplet injection affects cell viability. A droplet of inert silicone oil, corresponding in size to the photoresist droplets, was injected into each cell. In all control experiments, the cells were exposed to the same environmental conditions as the cells in which the structures were printed. Error bars represent standard error.
}
	\label{fig:2}
\end{figure}

One of the crucial parts of this study was determining the effect of 3D printing on the cells, specifically the cell viability. Already by time-lapse imaging, it can be observed that viable cells containing the printed structures have normal morphology and cell dynamics (Supplementary Video 1). The cells were observed to divide, and the structure was passed to one of the daughter cells (Fig.\ \ref{fig:2}a and Supplementary Video 2). The printed structure shown in Fig.\ \ref{fig:2}a was in the shape of the J. Stefan Institute logo. Its computer-aided design (CAD) and scanning electron microscopy (SEM) image are shown in Figs. \ref{fig:2}b and c. Confocal images show that the internal structure of the cells deforms to accommodate the printed structures, which can be seen especially well for the nucleus (Fig.\ \ref{fig:2}d).

We quantified how micropipette membrane penetration, sudden cellular volume increase, and membrane reshaping due to the droplet injection, toxicity of the photoresist, and laser illumination, affect the viability of cells (Fig.\ \ref{fig:2}e). We checked how the membrane penetration influences viability by injecting a small volume of a dyed medium, commonly used in microinjection experiments. To check how the sudden volume increase affects the viability, inert silicone oil was injected into the cells, forming a droplet comparable in size to typical photoresist droplets. The effects of photoresist toxicity and illumination were tested simultaneously. 

Microinjection is a method standard for delivering foreign materials into individual cells. Membrane penetration is known to cause damage that sometimes leads to cell death. In literature, the reported viability figures for microinjection vary considerably, from \SIrange{5}{15}{\%} for manual microinjections into neurons  \cite{lappe2008microinjection}, to the 96 \% for a semi-automated microrobotic microinjection into adherent cells \cite{wang2008system}. In our case, already in the control without injection, there were \SI{10}{\%}  non-viable cells after \SI{24}{h}, probably due to the prolonged handling of cells at ambient conditions. By comparing the FITC-dextran injected cells to the control, we can see that the microinjection causes a large increase of non-viable cells to \SI{50}{\%}. This percentage is similar also for the silicone-oil-injected cells (\SI{44}{\%}) and for cells with structures printed inside (\SI{55}{\%}). This suggests that the membrane penetration, which these three cases have in common, is the main cause of cells dying. However, there are probably additional smaller contributions: a mechanical effect due to introducing a large object into the cell, and an effect of printing, either due to the photoresist toxicity, illumination, or both. By comparing the number of cells that were flushed away during the exchange of the culture medium, we can see that this is not connected to the membrane penetration, since the percentages of flushed cells are almost identical for control and FITC-dextran injected cells. However, the flushing is more pronounced in cells injected with silicone oil and photoresist, which suggests that injecting a viscous droplet into the cell increases its probability of being flushed away. After the injection, the membrane needs to reshape considerably to accommodate the droplet, which probably weakens the cell's attachment to the substrate, making it more prone to completely detach if a strong flow appears during pipetting.

At this proof-of-concept stage, our study was not optimized to maximize cell viability. In the future, the overall viability of the cells could be significantly improved by using biocompatible photoresists explicitly developed for printing inside cells, optimizing the injection procedure, as well as cell handling and printing at physiological conditions inside a cell incubator.

\section*{Quality of printed structures}
\begin{figure}[t!]
	\centering
	\includegraphics[width=17cm]{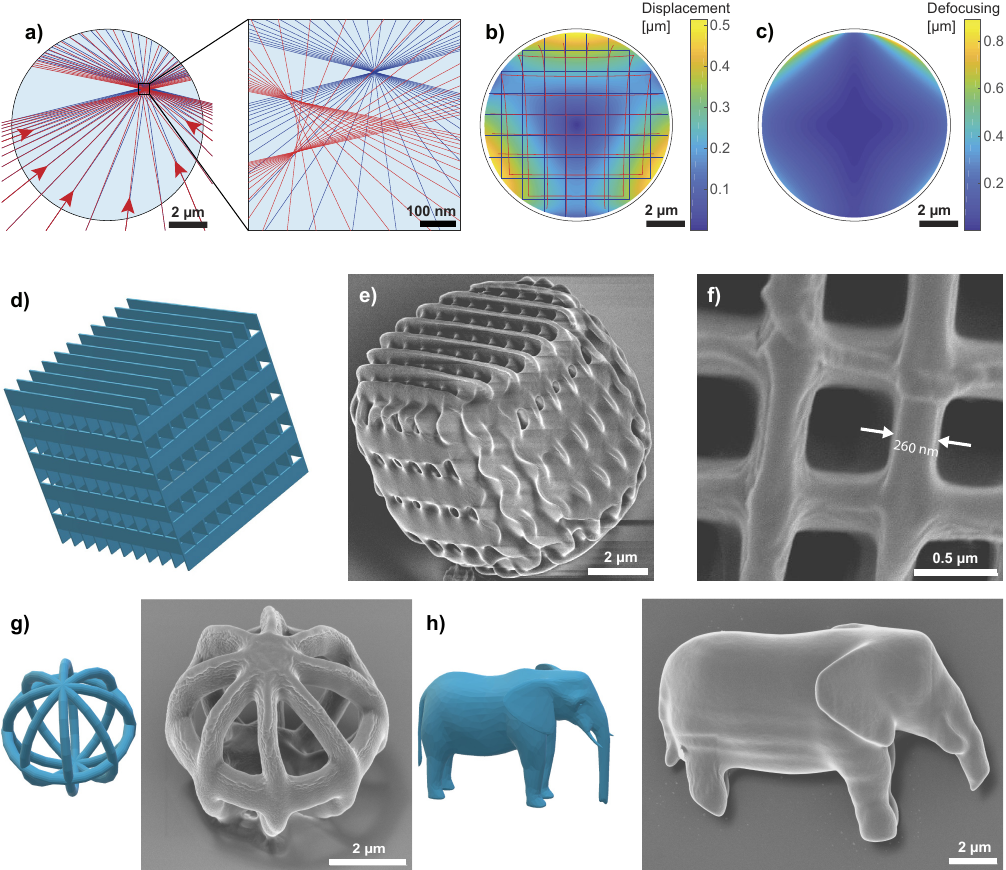}
	\caption{Printing quality analysis. (a) Focused beams of light with (red) or without (blue) the droplet present. The beams are drawn in one plane only. Due to the refraction on the droplet surface, the beams are not focused to a single point, and the overall focal point is shifted, as seen in the close-up on the right. (b) A colorplot, representing focal displacements in a meridian plane, as obtained in a 3D simulation, with a deformation of a square grid plotted for clarity. (c) A colorplot, representing the focal spot size in a meridian plane as obtained in a 3D simulation. (d) A CAD design of a woodpile structure. (e) A SEM image of the woodpile structure printed in a photoresist droplet. (f) A top-view SEM image of a grid in a structure, made by the same CAD design. (g) A CAD design and a SEM image of a \SI{6}{\micro m} hollow spherical structure printed in a droplet. (h) A CAD design and a SEM image of an elephant printed in a droplet (model downloaded from Tinkercad Shapes Library, designed by the user "geometricity"). 
}
	\label{fig:3}
\end{figure}

In TPP lithography, the structures are usually printed in a bulk photoresist. In our case, the printing occurs in a droplet of photoresist within a cell, therefore experiencing a substantial refractive-index mismatch. The refractive index of the unpolymerized photoresist is 1.48 and that of the cytoplasm is between 1.36 and 1.39 \cite{liu2016cell}. The refraction of the laser light through a curved droplet surface could result in distortions of the structures being printed and a decreased resolution. When viewed under an optical microscope, the structures printed in cells do not appear deformed. Only when the structure is larger than the droplet, the printed object can be truncated at the edge of the droplet (Fig.\ \ref{fig:2}a,c).

We performed a ray-optics simulation, where we analyzed how printing inside a droplet with a different refractive index than the surrounding medium affects the shape of the printed structures. Due to refraction at the droplet surface, the focus is shifted and the focal spot size increased (Fig.\ \ref{fig:3}a). This image displays the difference between the non-refracted beams (blue) and the beams refracted at the droplet surface (red) when focusing to a point within the droplet. For better visualization, the beams are drawn for a 2-dimensional case, whereas all the quantitative results were calculated in 3 dimensions. We have found that the displacements of the focus are small and amount to a maximum of \SI{0.5}{\micro m} near the droplet edge (Fig.\ \ref{fig:3}b). The image represents displacements in a selected plane intersecting the poles of the droplet; however, due to the rotational symmetry of the problem, the results are equal for all meridian planes. The deformation is already very small, but if needed, it could be easily completely eliminated by pre-deforming the model in the opposite direction to the expected shifts in the droplet due to diffraction.

The defocusing, shown in a meridian plane (Fig.\ \ref{fig:3}c), is below the diffraction limit of \SI{400}{nm} in more than 90 \% of the droplet volume. This means that the structures can be printed inside droplets with a resolution very close to the one achievable in bulk. Additional results from the simulation are presented in Supplementary Fig.\ 4. 

To experimentally confirm that there are no significant distortions or loss of resolution of the structures printed within droplets, we took SEM images of the structures printed in \SIrange{10}{15}{\micro m} droplets. In this case, the droplets were not in cells but in a mixture of glycerol and water, which had a refractive index close to that of the cytoplasm. We printed a woodpile structure, built from walls of single-voxel width, to evaluate the achievable printing resolution (Fig.\ \ref{fig:3}d). The design (Fig.\ \ref{fig:3}e) was slightly larger than the droplet. Therefore, the printed structure is slightly rounded on the sides. However, the grid on top appears flat and does not contain visible deformations, confirming that the in-droplet printing yields structures of high fidelity. The top view of the grid of another structure printed by the same design (Fig.\ \ref{fig:3}f) further confirms a good printing quality, with the grid exhibiting walls as thin as \SI{\sim 260}{nm} with a periodicity of \SI{\sim 0.8}{\micro m}. In this study, the printing parameters were not specifically optimized to achieve the highest resolution. However, the achieved 260-nm feature size is not far from the one achievable in bulk. SEM images of other designs, printed in droplets, are shown in Figs.\ \ref{fig:3}g, h, and in Supplementary Fig.\ 5. This demonstrates that any 3D structure can be printed with high resolution and fidelity, including structures with voids (Fig.\ \ref{fig:3}g).

\section*{Applications of the printed structures}

\begin{figure}[t!]
	\centering
	\includegraphics[width=17cm]{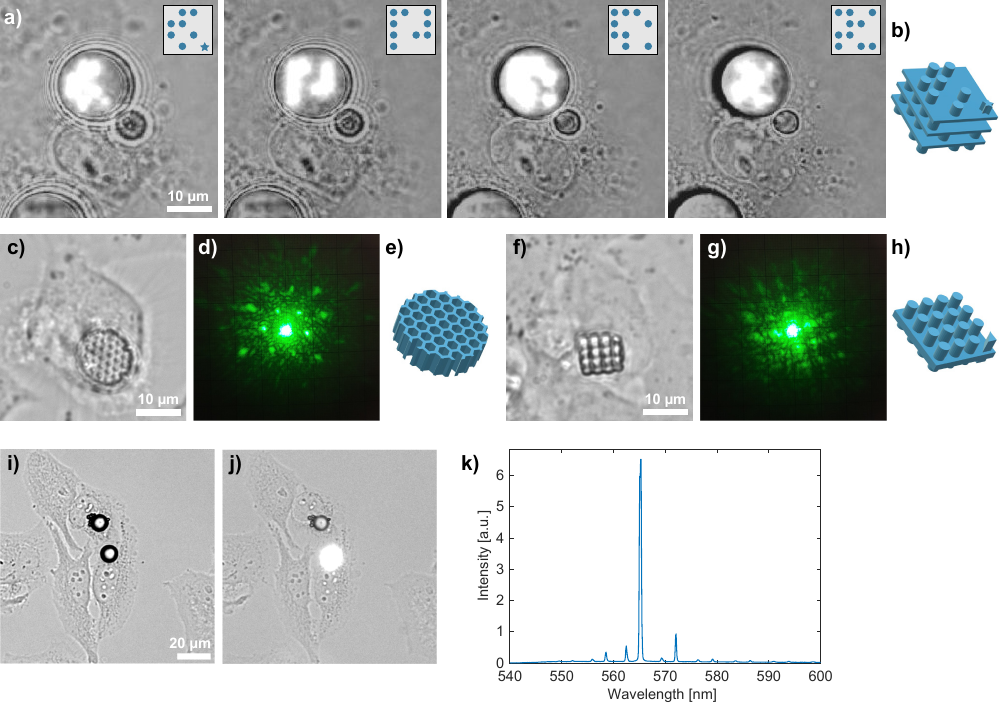}
	\caption{Applications. (a) Layer-by-layer printing of a 3D barcode inside a photoresist droplet within a HeLa cell. Insets show designs of the corresponding layer. (b) 3D design of the 3D barcode.
    (c) A hexagonal diffraction grating printed inside a HeLa cell, (d) its diffraction pattern, and (e) the 3D design. (f) A square diffraction grating printed inside a HeLa cell, (g) its diffraction pattern, and (h) the 3D design. (i) Two WGM lasers were made by single-photon polymerization of a droplet of high refractive index photoresist IP-n162, each injected into its separate HeLa cell. (j) Laser emission from a WGM laser within a cell, and (k) the corresponding emission spectrum. 
}
	\label{fig:4}
\end{figure}

There are many promising applications of printing within living cells. One of the applications we explored is barcoding, which involves writing a specific code to each cell for the purpose of identification and long-term tracking of the cell. We have designed a 3D graphical barcode composed of 4 stacked 4x4 grids of cylinders (Fig.\ \ref{fig:4}a, Supplementary Fig.\ 5c, and Supplementary Video 3). Each of the 64 spaces can be occupied or not. In the middle two layers, at least one cylinder should be present so that the structure does not fall apart. One cylinder on the top corner is also always present and is star-shaped so that the orientation of the structure can be determined. Therefore, we can encode 61 bits of information, giving a total of ($2 \cdot 10^{18}$) unique barcodes. This number far exceeds the number of cells in the human body. For real-life applications, much smaller barcodes could be used. Compared to most other cell barcoding techniques \cite{martino2019wavelength,klein2015droplet}, our method uniquely allows for encoding predefined information into each cell, rather than relying on random barcodes.

For remote reading, diffraction gratings of different designs can also be used as cell barcodes (Figs.\ \ref{fig:4}c-h). By illuminating them with a CW laser, a diffraction pattern that depends on the symmetry and periodicity of the structure can be read from a distance. Additionally, as the diffraction pattern is orientation-dependent, a diffraction grating could also be used to remotely measure cellular rotations.

We have further explored whether we can manufacture an active optical device, specifically a microlaser, inside a cell \cite{schubert2015lasing,humar2015intracellular}. A microsphere with a refractive index higher than the surrounding, can act as a whispering gallery mode (WGM) resonator. If it contains a gain medium, it can support WGM lasing when pumped above the threshold. By injecting fluorescently doped high-refractive-index photoresist IP-n162 into live cells and polymerizing the whole droplets, we managed to produce WGM lasers with diameters as small as \SI{9}{\micro m} (Figs.\  \ref{fig:4}i-k and Supplementary Fig.\ 6). While WGM lasers in the form of solid fluorescent beads or semiconductor nanodisks, uptaken into the cell through phagocytosis, have been used for cell tagging and intracellular sensing before \cite{humar2017spectral, martino2019wavelength, kavcic2022deep, schubert2015lasing}, the advantage of our approach is that the laser can be produced in-situ, which enables its use as a barcode in cells that cannot uptake larger particles. The high-refractive-index photoresist that we used appears more cytotoxic than IP-S, causing all of the injected cells to either die (80 \%) during the course of 22 hours, or at least be visibly affected. However, with the constant development of new bio-compatible photo-curable materials, we expect this setback can be overcome.

Instead of using photoresists only slightly soluble in water, it is also possible to inject a water-soluble hydrogel-based photoresist into the cells. In this case, the photoresist gets distributed throughout the cell, enabling printing anywhere within, thus making it possible to isolate a certain part of the cell inside a newly made compartment, for example, to study the individual functioning of certain organelles or to investigate signaling pathways by physically blocking some of them. In this approach, instead of inactivating or physically destroying organelles to study their regeneration or biogenesis \cite{jollivet2007analysis,tangemo2011novel}, it would be possible to fix them or entire selected cells by photo-polymerization. This would enable studying how mechanical properties affect different cellular behaviors or the modeling of diseases that involve changes in tissue stiffness. Preliminary results of this approach are presented in the Supplementary Note and Supplementary Fig.\ 7.

\section*{Conclusions}
For the first time, we printed various polymeric structures directly inside living cells. While microstructures can be introduced into cells by phagocytosis, as shown before  \cite{gomez2013silicon,schubert2017lasing}, only a small number of cell types have phagocytotic capacity. Also, in this case, the structure ends up in a phagosome and is not directly inside the cytosol. Therefore, printing the devices within cells overcomes these limitations.

We have developed a printing protocol and found a commercially available photoresist with which such printing is possible. Contrary to the usual in-bulk printing, in this work, 3D printing took place inside a \SI{\sim 10}{\micro m} droplet of a photoresist, which has previously been demonstrated only in much larger droplets \cite{afting2024minimal}. We performed simulations to characterize this printing and found that the deformations arising from the refractive index mismatch are small and allow for very detailed structures with a sub-micron resolution. Many cells seemed unperturbed by the printing procedure and also underwent cellular division. Our study is only a proof of concept, and can in the future be further optimized to achieve better viability.

There are many possible applications of structures printed inside the cells, well beyond what has been shown here. Especially interesting is the prospect of printing functional structures, which would change the properties of cells beyond what has been possible till now by conventional bioengineering tools. Structures, such as micro-levers, springs, cages, and barriers printed in cells, could be used to apply controlled mechanical forces to intracellular components, introduce chirality, and modify cell shape and cellular mechanics. Such modified cells could help us study cell mechanics, cellular division, mechano-transduction, and mechanically induced differentiation of cells in culture and in tissues.

Here, we printed diffraction gratings and lasers, but other optical devices, such as lenses \cite{chen2021lipid}  and waveguides for controlled light delivery and imaging, could also be used.

With the fast development of new functional materials for photolithography, TPP could be used to fabricate various active devices inside living cells. Functional micro-devices are produced by 4D TPP printing \cite{jian2023two, wan2024recent, mainik2024recent} using different functional materials\cite{carlotti2019functional}, responsive to various external stimuli, such as light \cite{jung2020designing, hippler2019controlling}, temperature \cite{spiegel20224d, zhang2021structural}, pH \cite{hu2020botanical}, sugar concentration \cite{ennis2023two}, and magnetic field \cite{williams2018two}. This would enable both novel sensing modalities and the creation of active components and even microrobots inside cells. Further, TPP can also be performed with conductive materials \cite{blasco2016fabrication, lichade2024direct}, even inside living organisms \cite{baldock2023creating}. Conductive micro-structures printed in live cells could lead to new intracellular electrophysiology modalities \cite{xu2021cell, liu2020heart}. By using nanodiamond-embedded photoresists, even quantum sensing with a 3D printed quantum sensor \cite{blankenship2023complex} might be possible inside cells. Further, printing a structure loaded with a drug \cite{yan2014micro} with controlled size, shape, and position within the cell could enable spatio-temporally controlled intracellular drug release.

Intracellular 3D printing offers an unprecedented degree of control over the cellular interior, allowing the integration of synthetic structures with native biological functions. This platform could allow for reconfiguration of cellular architecture, embed logic or mechanical components within the cytoplasm, and design cells with enhanced or entirely new properties. These developments could have far-reaching implications for synthetic biology, intracellular sensing, and the fundamental study of cellular structure and function.

\section*{Materials and methods}

\subsection*{Materials for injection}

Materials used for injection into live cells: IP-S photoresist (Nanoscribe GmbH) with a refractive index 1.486 in liquid form and 1.515 in polymerized form, IP-n162 photoresist (Nanoscribe GmbH) with a refractive index of 1.604 in liquid form and 1.622 in polymerized form, dyed with Pyrromethene 597 (Exciton, USA), and Tris-EDTA (TE) buffer dyed with \SI{1}{\milli M} FITC-dextran (\SI{10}{kDa}, Sigma-Aldrich). For silicone oil injection two inert silicon oils were used yielding similar results, namely Antievaporation oil (Ibidi, Germany) and
Poly(dimethylsiloxane-co-methylphenylsiloxane) (Dow Corning, USA).

\subsection*{Cell culture}
HeLa cells (ATCC, CCL-2) were seeded into a well, consisting of a silicone layer with a \SI{10}{mm} circular cut-out (Silicone Isolators, Grace Bio-labs) on a previously cleaned coverslip, 1-2 days before the printing. Cells were grown overnight in the cell-culture medium DMEM (Dulbecco's Modified Eagle Medium, DMEM (1X) + GlutaMAX, Gibco) supplemented with 10 \% fetal bovine serum and 1 \% penicillin-streptomycin in an incubator (\SI{37}{^{\circ} C}, 5 \% CO$_2$) to achieve approximately 70 \% confluence. Before cell injection, the cell growth medium was exchanged for LCIS (Live Cell Imaging Solution, Molecular Probes, Invitrogen) supplemented with \SI{10}{mM} glucose. The LCIS was used because the injection and printing took place at room conditions.

\subsection*{Micropipette preparation}
The micropipettes were pulled on the P1000 Micropipette Puller (Sutter Instruments) from borosilicate glass capillaries with a filament (outer diameter \SI{1}{mm}, inner diameter \SI{0.78}{mm}, BF100-78-10, Sutter Instruments) using the following program settings: heat 467, pull 90, velocity 70, delay 90, pressure 220, resulting in a micropipette tip with a sub-micron outer diameter (Supplementary Fig.\ 1a).
On the day of the injections, the previously pulled micropipettes were silanized to facilitate wetting by the photoresist. For 3 minutes, the inside of the micropipette was exposed to a vapour of silane (Dimethyloctadecyl[3-(trimethoxysilyl)propyl]ammonium chloride solution, Aldrich, Germany), produced by heating the solution to \SI{60}{^{\circ} C}. Afterwards, the silane was baked for 20 minutes at \SI{120}{^{\circ} C}. After silanization, the micropipette was filled with the photoresist. For injection of the TE buffer dyed with FITC-dextran, the silanization step was not performed.

\subsection*{Injection into cells}
Injection of photoresists and other materials into the cells was performed with a FemtoJet 4i microinjector (Eppendorf), coupled with an InjectMan 4 micromanipulator (Eppendorf). 
For injection, we randomly selected cells from the ones that appeared well attached to the substrate (well spread-out). Otherwise, they were very likely to detach during injection and stick to the micropipette. Microinjection was performed at room temperature, without the added CO$_2$, in LCIS. During one experiment, up to 40 droplets were injected. Typical injection pressure $p_i$ and injection time $t_i$ for IP-S photoresist were $p_i =\SI{5000}{hPa}$, $t_i =\SI{25}{s}$, for silicon oil $p_i =\SI{4000}{hPa}$, $t_i =\SI{1.7}{s}$, for TE buffer dyed with FITC-dextran $p_i =\SI{1000}{hPa}$, $t_i =\SI{4}{s}$, and for IP-n162 photoresist $p_i =\SI{5000}{hPa}$, $t_i =\SIrange{40}{90}{s}$.

\subsection*{3D-printing protocol}
After taking the sample out of the incubator,  DMEM was exchanged for LCIS. This was followed by the microinjection step, which took approximately 30 minutes, and then the printing procedure commenced. A commercial system, Photonic Professional GT2 (Nanoscribe, Germany), was used for printing. The printing step lasted for approximately 45 minutes, and was performed at ambient conditions. Most of this time was used to mount (and later dismount) the sample and to manually search and appropriately position each droplet into the location of the laser spot. After positioning, each droplet was illuminated with a \SI{780}{nm} femtosecond laser through a high numerical aperture 63x objective (Zeiss - 1.4 NA Oil DIC, Plan Polychromat). The system operated in the galvo mode, with the scan speed of \SI{10 000}{\micro m} s$^{-1}$. Illumination of a single structure took \SIrange{3}{10}{s}, depending on the structure volume. After printing all the structures, the LCIS medium was exchanged for DMEM, and the sample was put into an incubator. The typical experimental timeline is shown in Supplementary Fig.\ 2. 

The codes for printing different 3D structures were prepared in the proprietary Describe software, where the CAD design was automatically sliced into \SI{100}{nm} layers, which would be illuminated one after the other. Within each plane, the laser was programmed to fill the structure by moving in parallel lines, separated by \SI{100}{nm}. The structures were coded to be printed top-down, so first the photoresist furthest away from the glass was to be illuminated and then layers closer and closer to the glass. Due to the high viscosity of the photoresist and short printing times, the subsidence of structures during printing was not problematic.
 
\subsection*{Viability study}
 To check the extent to which the microinjection damages the cells, 2 different sets of control experiments were performed. In one, a small amount of microinjection buffer (Tris-EDTA (TE) buffer, typically used in DNA microinjection experiments) dyed with FITC-dextran was injected into the cells to see how damaging the puncturing through the membrane is by itself. The dye was added to find the microinjected cells next day, after the incubation. In another control experiment, inert silicone oil was injected into the cells. This was physically much more similar to the injection of the photoresist, as the silicone oil formed similarly sized droplets within the cells. Viability was assessed by performing a live/dead assay 24 hours after the injection. We compared these results to the viability of the cells that were subjected to the same environmental conditions but were not mechanically manipulated in any way. Microinjection and printing were performed at room conditions, with cells submerged in a medium designed for room-condition imaging. The timeline of changes of the outside conditions (temperature, medium exchanges, and sample put into the incubator) was kept the same in all experiments. The first 75 minutes, the samples were kept at room conditions in the LCIS, then, the LCIS was exchanged for the cell-growth medium DMEM, containing \SI{20}{nM} Sytox Green Nucleic Acid Stain (Invitrogen) or \SI{50}{nM} Sytox Deep Red Nucleic Acid Stain (Invitrogen). The samples were imaged and put into the incubator. At 24 hours, the samples were checked again to see how many of the observed cells were still viable. The bar chart in Fig.\ \ref{fig:2}(a) shows combined results obtained from multiple repetitions of the experiment. Each repetition was performed on a new sample of cells.

\subsection*{Conventional and confocal microscopy}
Brightfield imaging was performed on an inverted microscope (Nikon Ti2). The samples were placed into a tabletop incubator. To reduce the photo-toxicity, the white LED illumination was set to a low value and filtered with a \SI{600}{nm} longpass filter. Imaging was performed with 10X, 20X, and 40X objectives with a monochrome camera (Ximea xiC MC203MG-SY-UB). 

For viability studies, a bright-field image was taken after the photoresist injection, after printing and medium exchange, and at the end of the 24-hour experiment. After medium exchange and at the end of the experiment, a fluorescence image, where the nuclei of all the nonviable cells were labelled green or red, depending on the Sytox dye used, was taken in addition to the bright-field image. In some experimental repetitions, time-lapse imaging was performed by taking a bright-field image every 10 minutes (or every 5 minutes) during the 24-hour experiment. Additionally, every few hours and at the end of the experiment, a fluorescence image was taken. Counting of viable and nonviable cells was done manually.

For the time-lapse imaging of the cells in the fluorescence mode (Supplementary Video 1), the cell nuclei were labeled with NucSpot Live 488 (by Biotium, USA) and the cell membranes with CellMask Orange dye (by Invitrogen, USA). A time-lapse image sequence was recorded by a color camera (IDS, UI-3280CP-C-HQ R2). 

For confocal imaging, the samples were fixed by immersion in 1 \% paraformaldehyde (PFA, Biotium, USA) for 5 minutes, and a subsequent immersion in 2 \% PFA for 10 minutes, washed with phosphate-buffered saline (PBS, Gibco) and labelled with CellMask Deep Red (Invitrogen, USA) for \SI{10}{min} and NucSpot Live 488 (Biotium, USA) for \SI{30}{min} at room temperature. The IP-S structures did not need to be additionally labeled as the photoresist is fluorescent by itself. Confocal imaging was performed on a customized confocal microscope (Abberior). The grayscale images from separate channels were artificially colored and combined. Brightness and contrast were adjusted separately for each channel.

\subsection*{Scanning electron microscopy (SEM)}
For SEM imaging, the structures were printed in IP-S photoresist droplets dispersed in a glycerol solution in deionized water (1:3 volume ratio) in a \SI{50}{\micro l} well. This solution has a refractive index of 1.365. After printing, the well was submerged in \SI{5}{ml} deionized water for a couple of hours for the unpolymerized photoresist to dissolve. Afterwards, the samples were dried and coated with an approximately \SI{10}{nm} Au/Pd layer using a Gatan 682 Precision Etching and Coating System (PECS). SEM imaging was performed using a FEI Helios Nanolab 650 scanning electron microscope. High resolution/high magnification imaging was performed in immersion mode with \SI{2}{keV} electron acceleration voltage, \SI{25}{pA} electron current, and $10^{-6}$ \SI{}{hPa} chamber vacuum. For better 3D visualization of the 3D-printed objects, the stage was tilted up to 50 degrees.

\subsection*{Microlaser operation and analysis}

A WGM microlaser consisted of a high-refractive-index photoresist dyed with a fluorescent dye (Pyrromethene 597) to act as a gain medium. After injection of the photoresist droplet into the cell, the whole droplet was shortly (\SI{5}{s}) illuminated with a UV LED light to polymerize (one-photon process). Then, the droplet was pumped with a tunable pulsed nanosecond laser (NT242, Ekspla, Lithuania) with a wavelength set to \SI{532}{nm} and repetition rate of \SI{10}{Hz}. The emission spectrum was measured with a spectrometer (Andor Shamrock, SR-500i-D1) equipped with a CCD camera (Newton, DU940P-BV), with a spectral resolution of \SI{0.13}{nm}. To characterize the laser, a model system was used, with a photoresist/dye droplet immersed in a water environment instead of in a cell. A typical two-slope curve was obtained when increasing the pump laser pulse energy and measuring the intensity of one of the spectral peaks, indicating a lasing threshold of approximately \SI{0.07}{nJ \micro m^{-2}} (Supplementary Fig.\ 6). A decrease of the linewidth was also observed at this value.

\subsection*{Simulations}
The precision of the printing process in terms of displacement and width of the beam at different positions inside the droplet was calculated by simulating the propagation of rays. A sufficient number ($\sim 2500$, see Supplementary Fig.\ 4b) of rays with their origins and angles of propagation were selected. The origins were homogeneously distributed in a plane below the droplet, while the angles were such that all the rays would converge to the same focal position in the absence of refraction. The maximum entrance angle was calculated based on the numerical aperture of the objective and the refractive indices of the materials used in the experiment. Due to refraction at the curved surface of the droplet, different rays refract differently, leading to a wider and displaced area of ray intersections. The new focal position was assigned by looking at the distribution of rays in each vertical plane. The plane where the variance of ray positions was the smallest was defined as the z position of the focus, while x and y positions were taken as the center of mass in that plane. Displacement was then calculated as the difference between this position and the position if no refraction would take place. The focus width or resolution was defined as the variance of ray positions in the newly-assigned focal plane. In the simulation, the ratio between the refractive indices of unpolymerized photoresist IP-S and cytoplasm was 1.08. The droplet diameter was \SI{10}{\micro m}. 
For the calculation of the focus displacement and defocusing, the simulation was ran only for the volume within the 0.96 of the droplet radius (Fig.\ \ref{fig:3}b,c) as at the upper edges of the droplet the rays are refracted in a manner that they can become parallel and do not intersect at all, resulting in the divergence of the simulation. For calculating the percentage of volume, where defocusing was below the diffraction limit, the simulation additionally included radii up to the full droplet radius, and the diverging points were attributed a defocusing value above the diffraction limit.

\section*{Data availability}
The data that support the findings of this study are available from the corresponding author upon reasonable request.

\section*{Code availability}
The code that supports the findings of this study is available from the corresponding
author upon reasonable request.

\section*{Acknowledgments}

We acknowledge Ana Krišelj for preparing cell samples, Boštjan Kokot for taking confocal images, and Maja Zorc for the initial photoresist cytotoxicity experiments. This project has received funding from the European Research Council (ERC) under the European Union’s Horizon 2020 research and innovation programme (grant agreement No. 851143) and from the Slovenian Research and Innovation Agency (ARIS) (N1-0362 and P1-0099).

\section*{Conflict of Interest}
The authors declare no conflict of interest.

\section*{Author Contributions}
M. M. conducted experiments and analyzed the results. A. K. performed simulations and analyzed the results. In initial experiments, U. J. performed the 3D-printing part of the experiment and later provided valuable insight regarding the 3D printing. R. P. prepared the samples for SEM imaging and took SEM images. M.H. conceived the original idea and designed and supervised the study. M. M. and M. H. wrote the manuscript with input from the other authors. All authors approved the final version of the paper.

 \typeout{} 
 \bibliography{Bibliography}

\begin{thebibliography}{10}

\bibitem{espera20193d}
Alejandro~H Espera, John Ryan~C Dizon, Qiyi Chen, and Rigoberto~C Advincula.
\newblock {3D}-printing and advanced manufacturing for electronics.
\newblock {\em Progress in Additive Manufacturing}, 4:245--267, 2019.

\bibitem{wallin20183d}
TJ~Wallin, James Pikul, and Robert~F Shepherd.
\newblock {3D} printing of soft robotic systems.
\newblock {\em Nature Reviews Materials}, 3(6):84--100, 2018.

\bibitem{gonzalez2023micro}
Diana Gonzalez-Hernandez, Simonas Varapnickas, Andrea Bertoncini, Carlo
  Liberale, and Mangirdas Malinauskas.
\newblock Micro-optics {3D} printed via multi-photon laser lithography.
\newblock {\em Advanced Optical Materials}, 11(1):2201701, 2023.

\bibitem{wang2023two}
Hao Wang, Wang Zhang, Dimitra Ladika, Haoyi Yu, Darius Gailevi{\v{c}}ius,
  Hongtao Wang, Cheng-Feng Pan, Parvathi Nair~Suseela Nair, Yujie Ke, Tomohiro
  Mori, et~al.
\newblock Two-photon polymerization lithography for optics and photonics:
  fundamentals, materials, technologies, and applications.
\newblock {\em Advanced Functional Materials}, 33(39):2214211, 2023.

\bibitem{hippler20193d}
Marc Hippler, Enrico~Domenico Lemma, Sarah Bertels, Eva Blasco, Christopher
  Barner-Kowollik, Martin Wegener, and Martin Bastmeyer.
\newblock {3D} scaffolds to study basic cell biology.
\newblock {\em Advanced Materials}, 31(26):1808110, 2019.

\bibitem{guvendiren20193d}
Murat Guvendiren.
\newblock {\em {3D} bioprinting in medicine: technologies, bioinks, and
  applications}.
\newblock Springer, 2019.

\bibitem{maruo1997three}
Shoji Maruo, Osamu Nakamura, and Satoshi Kawata.
\newblock Three-dimensional microfabrication with two-photon-absorbed
  photopolymerization.
\newblock {\em Optics Letters}, 22(2):132--134, 1997.

\bibitem{liao2020material}
Caizhi Liao, Alain Wuethrich, and Matt Trau.
\newblock A material odyssey for {3D} nano/microstructures: two photon
  polymerization based nanolithography in bioapplications.
\newblock {\em Applied Materials Today}, 19:100635, 2020.

\bibitem{huang2021highly}
Xing Huang, Yuxi Zhang, Mengquan Shi, Li-Peng Zhang, Yunlong Zhang, and Yuxia
  Zhao.
\newblock A highly biocompatible bio-ink for {3D} hydrogel scaffolds
  fabrication in the presence of living cells by two-photon polymerization.
\newblock {\em European Polymer Journal}, 153:110505, 2021.

\bibitem{hasselmann2018attachment}
Nils~Frederik Hasselmann and Wolfgang Horn.
\newblock Attachment of microstructures to single bacteria by two-photon
  patterning of a protein based hydrogel.
\newblock {\em Biomedical Physics \& Engineering Express}, 4(3):035004, 2018.

\bibitem{urciuolo2020intravital}
Anna Urciuolo, Ilaria Poli, Luca Brandolino, Paolo Raffa, Valentina Scattolini,
  Cecilia Laterza, Giovanni~G Giobbe, Elisa Zambaiti, Giulia Selmin, Michael
  Magnussen, et~al.
\newblock Intravital three-dimensional bioprinting.
\newblock {\em Nature Biomedical Engineering}, 4(9):901--915, 2020.

\bibitem{afting2024minimal}
Cassian Afting, Philipp Mainik, Clara Vazquez-Martel, Tobias Abele, Verena
  Kaul, Girish Kale, Kerstin G{\"o}pfrich, Steffen Lemke, Eva Blasco, and
  Joachim Wittbrodt.
\newblock Minimal-invasive {3D} laser printing of microimplants in organismo.
\newblock {\em Advanced Science}, page 2401110, 2024.

\bibitem{abele2022two}
Tobias Abele, Tobias Messer, Kevin Jahnke, Marc Hippler, Martin Bastmeyer,
  Martin Wegener, and Kerstin G{\"o}pfrich.
\newblock Two-photon {3D} laser printing inside synthetic cells.
\newblock {\em Advanced Materials}, 34(6):2106709, 2022.

\bibitem{arbore2019probing}
Claudia Arbore, Laura Perego, Marios Sergides, and Marco Capitanio.
\newblock Probing force in living cells with optical tweezers: from
  single-molecule mechanics to cell mechanotransduction.
\newblock {\em Biophysical Reviews}, 11(5):765--782, 2019.

\bibitem{timonen2017tweezing}
Jaakko~VI Timonen and Bartosz~A Grzybowski.
\newblock Tweezing of magnetic and non-magnetic objects with magnetic fields.
\newblock {\em Advanced Materials}, 29(18):1603516, 2017.

\bibitem{schubert2020monitoring}
Marcel Schubert, Lewis Woolfson, Isla~RM Barnard, Amy~M Dorward, Becky
  Casement, Andrew Morton, Gavin~B Robertson, Paul~L Appleton, Gareth~B Miles,
  Carl~S Tucker, et~al.
\newblock Monitoring contractility in cardiac tissue with cellular resolution
  using biointegrated microlasers.
\newblock {\em Nature Photonics}, 14(7):452--458, 2020.

\bibitem{kavcic2022deep}
Alja{\v{z}} Kav{\v{c}}i{\v{c}}, Maja Garvas, Matev{\v{z}} Marin{\v{c}}i{\v{c}},
  Katrin Unger, Anna~Maria Coclite, Boris Majaron, and Matja{\v{z}} Humar.
\newblock Deep tissue localization and sensing using optical microcavity
  probes.
\newblock {\em Nature Communications}, 13(1):1269, 2022.

\bibitem{gomez2013silicon}
Rodrigo G{\'o}mez-Mart{\'\i}nez, Alberto~M Hern{\'a}ndez-Pinto, Marta Duch,
  Patricia V{\'a}zquez, Kirill Zinoviev, Enrique~J de~La~Rosa, Jaume Esteve,
  Teresa Su{\'a}rez, and Jos{\'e}~A Plaza.
\newblock Silicon chips detect intracellular pressure changes in living cells.
\newblock {\em Nature Nanotechnology}, 8(7):517--521, 2013.

\bibitem{fernandez2009intracellular}
Elisabet Fernandez-Rosas, Rodrigo Gomez, Elena Ibanez, Leonardo Barrios, Marta
  Duch, Jaume Esteve, Carme Nogu{\'e}s, and Jos{\'e}~Antonio Plaza.
\newblock Intracellular polysilicon barcodes for cell tracking.
\newblock {\em Small}, 5(21):2433--2439, 2009.

\bibitem{humar2015intracellular}
Matja{\v{z}} Humar and Seok~Hyun Yun.
\newblock Intracellular microlasers.
\newblock {\em Nature Photonics}, 9(9):572, 2015.

\bibitem{martino2019wavelength}
Nicola Martino, Sheldon~JJ Kwok, Andreas~C Liapis, Sarah Forward, Hoon Jang,
  Hwi-Min Kim, Sarah~J Wu, Jiamin Wu, Paul~H Dannenberg, Sun-Joo Jang, et~al.
\newblock Wavelength-encoded laser particles for massively multiplexed cell
  tagging.
\newblock {\em Nature Photonics}, 13(10):720--727, 2019.

\bibitem{anwar2023microcavity}
Abdur~Rehman Anwar, Maru\v Mur, and Matja\v z Humar.
\newblock Microcavity-and microlaser-based optical barcoding: A review of
  encoding techniques and applications.
\newblock {\em ACS Photonics}, 10(5):1202--1224, 2023.

\bibitem{lappe2008microinjection}
Corinna Lappe-Siefke, Christoph Maas, and Matthias Kneussel.
\newblock Microinjection into cultured hippocampal neurons: A straightforward
  approach for controlled cellular delivery of nucleic acids, peptides and
  antibodies.
\newblock {\em Journal of Neuroscience Methods}, 175(1):88--95, 2008.

\bibitem{wang2008system}
Wenhui Wang, Yu~Sun, Ming Zhang, Robin Anderson, Lowell Langille, and Warren
  Chan.
\newblock A system for high-speed microinjection of adherent cells.
\newblock {\em Review of Scientific Instruments}, 79(10), 2008.

\bibitem{liu2016cell}
Patricia~Yang Liu, Lip~Ket Chin, Wee Ser, HF~Chen, C-M Hsieh, C-H Lee, K-B
  Sung, TC~Ayi, PH~Yap, Bo~Liedberg, et~al.
\newblock Cell refractive index for cell biology and disease diagnosis: past,
  present and future.
\newblock {\em Lab on a Chip}, 16(4):634--644, 2016.

\bibitem{klein2015droplet}
Allon~M Klein, Linas Mazutis, Ilke Akartuna, Naren Tallapragada, Adrian Veres,
  Victor Li, Leonid Peshkin, David~A Weitz, and Marc~W Kirschner.
\newblock Droplet barcoding for single-cell transcriptomics applied to
  embryonic stem cells.
\newblock {\em Cell}, 161(5):1187--1201, 2015.

\bibitem{schubert2015lasing}
Marcel Schubert, Anja Steude, Philipp Liehm, Nils~M Kronenberg, Markus Karl,
  Elaine~C Campbell, Simon~J Powis, and Malte~C Gather.
\newblock Lasing within live cells containing intracellular optical
  microresonators for barcode-type cell tagging and tracking.
\newblock {\em Nano Letters}, 15(8):5647--5652, 2015.

\bibitem{humar2017spectral}
Matja{\v{z}} Humar, Avinash Upadhya, and Seok~Hyun Yun.
\newblock Spectral reading of optical resonance-encoded cells in microfluidics.
\newblock {\em Lab on a Chip}, 17(16):2777--2784, 2017.

\bibitem{jollivet2007analysis}
Florence Jollivet, Gra{\c{c}}a Raposo, Ariane Dimitrov, Rachid Sougrat, Bruno
  Goud, and Franck Perez.
\newblock Analysis of de novo golgi complex formation after enzyme-based
  inactivation.
\newblock {\em Molecular Biology of the Cell}, 18(11):4637--4647, 2007.

\bibitem{tangemo2011novel}
Carolina T{\"a}ngemo, Paolo Ronchi, Julien Colombelli, Uta Haselmann, Jeremy~C
  Simpson, Claude Antony, Ernst~HK Stelzer, Rainer Pepperkok, and Emmanuel~G
  Reynaud.
\newblock A novel laser nanosurgery approach supports de novo golgi biogenesis
  in mammalian cells.
\newblock {\em Journal of Cell Science}, 124(6):978--987, 2011.

\bibitem{schubert2017lasing}
Marcel Schubert, Klara Volckaert, Markus Karl, Andrew Morton, Philipp Liehm,
  Gareth~B Miles, Simon~J Powis, and Malte~C Gather.
\newblock Lasing in live mitotic and non-phagocytic cells by efficient delivery
  of microresonators.
\newblock {\em Scientific Reports}, 7(1):40877, 2017.

\bibitem{chen2021lipid}
Xixi Chen, Tianli Wu, Zhiyong Gong, Jinghui Guo, Xiaoshuai Liu, Yao Zhang,
  Yuchao Li, Pietro Ferraro, and Baojun Li.
\newblock Lipid droplets as endogenous intracellular microlenses.
\newblock {\em Light: Science \& Applications}, 10(1):242, 2021.

\bibitem{jian2023two}
Bingcong Jian, Honggeng Li, Xiangnan He, Rong Wang, Hui~Ying Yang, and Qi~Ge.
\newblock Two-photon polymerization-based {4D} printing and its applications.
\newblock {\em International Journal of Extreme Manufacturing}, 6(1):012001,
  2023.

\bibitem{wan2024recent}
Xue Wan, Zhongmin Xiao, Yujia Tian, Mei Chen, Feng Liu, Dong Wang, Yong Liu,
  Paulo Jorge Da~Silva Bartolo, Chunze Yan, Yusheng Shi, et~al.
\newblock Recent advances in {4D} printing of advanced materials and structures
  for functional applications.
\newblock {\em Advanced Materials}, 36(34):2312263, 2024.

\bibitem{mainik2024recent}
Philipp Mainik, Christoph~A Spiegel, and Eva Blasco.
\newblock Recent advances in multi-photon {3D} laser printing: active materials
  and applications.
\newblock {\em Advanced Materials}, 36(11):2310100, 2024.

\bibitem{carlotti2019functional}
Marco Carlotti and Virgilio Mattoli.
\newblock Functional materials for two-photon polymerization in
  microfabrication.
\newblock {\em Small}, 15(40):1902687, 2019.

\bibitem{jung2020designing}
Kenward Jung, Nathaniel Corrigan, Mustafa Ciftci, Jiangtao Xu, Soyoung~E Seo,
  Craig~J Hawker, and Cyrille Boyer.
\newblock Designing with light: advanced {2D}, {3D}, and {4D} materials.
\newblock {\em Advanced Materials}, 32(18):1903850, 2020.

\bibitem{hippler2019controlling}
Marc Hippler, Eva Blasco, Jingyuan Qu, Motomu Tanaka, Christopher
  Barner-Kowollik, Martin Wegener, and Martin Bastmeyer.
\newblock Controlling the shape of {3D} microstructures by temperature and
  light.
\newblock {\em Nature Communications}, 10(1):232, 2019.

\bibitem{spiegel20224d}
Christoph~A Spiegel, Maximilian Hackner, Viktoria~P Bothe, Joachim~P Spatz, and
  Eva Blasco.
\newblock {4D} printing of shape memory polymers: from macro to micro.
\newblock {\em Advanced Functional Materials}, 32(51):2110580, 2022.

\bibitem{zhang2021structural}
Wang Zhang, Hao Wang, Hongtao Wang, John You~En Chan, Hailong Liu, Biao Zhang,
  Yuan-Fang Zhang, Komal Agarwal, Xiaolong Yang, Anupama~Sargur Ranganath,
  et~al.
\newblock Structural multi-colour invisible inks with submicron {4D} printing
  of shape memory polymers.
\newblock {\em Nature Communications}, 12(1):112, 2021.

\bibitem{hu2020botanical}
Yanlei Hu, Zhongyu Wang, Dongdong Jin, Chenchu Zhang, Rui Sun, Ziqin Li, Kai
  Hu, Jincheng Ni, Ze~Cai, Deng Pan, et~al.
\newblock Botanical-inspired {4D} printing of hydrogel at the microscale.
\newblock {\em Advanced Functional Materials}, 30(4):1907377, 2020.

\bibitem{ennis2023two}
Alexa Ennis, Deanna Nicdao, Srikanth Kolagatla, Luke Dowling, Yekaterina Tskhe,
  Alex~J Thompson, Daniel Trimble, Colm Delaney, and Larisa Florea.
\newblock Two-photon polymerization of sugar responsive {4D} microstructures.
\newblock {\em Advanced Functional Materials}, 33(39):2213947, 2023.

\bibitem{williams2018two}
Gwilym Williams, Matthew Hunt, Benedikt Boehm, Andrew May, Michael Taverne,
  Daniel Ho, Sean Giblin, Dan Read, John Rarity, Rolf Allenspach, et~al.
\newblock Two-photon lithography for {3D} magnetic nanostructure fabrication.
\newblock {\em Nano Research}, 11:845--854, 2018.

\bibitem{blasco2016fabrication}
Eva Blasco, Jonathan M{\"u}ller, Patrick M{\"u}ller, Vanessa Trouillet, Markus
  Sch{\"o}n, Torsten Scherer, Christopher Barner-Kowollik, and Martin Wegener.
\newblock Fabrication of conductive {3D} gold-containing microstructures via
  direct laser writing.
\newblock {\em Advanced Materials}, 28(18):3592--3595, 2016.

\bibitem{lichade2024direct}
Ketki~M Lichade, Shahrzad Shiravi, John~D Finan, and Yayue Pan.
\newblock Direct printing of conductive hydrogels using two-photon
  polymerization.
\newblock {\em Additive Manufacturing}, 84:104123, 2024.

\bibitem{baldock2023creating}
Sara~J Baldock, Punarja Kevin, Garry~R Harper, Rebecca Griffin, Hussein~H
  Genedy, M~James Fong, Zhiyi Zhao, Zijian Zhang, Yaochun Shen, Hungyen Lin,
  et~al.
\newblock Creating {3D} objects with integrated electronics via multiphoton
  fabrication in vitro and in vivo.
\newblock {\em Advanced Materials Technologies}, 8(11):2201274, 2023.

\bibitem{xu2021cell}
Dongxin Xu, Jingshan Mo, Xi~Xie, and Ning Hu.
\newblock In-cell nanoelectronics: opening the door to intracellular
  electrophysiology.
\newblock {\em Nano-Micro Letters}, 13(1):127, 2021.

\bibitem{liu2020heart}
Haitao Liu, Olurotimi~A Bolonduro, Ning Hu, Jie Ju, Akshita~A Rao, Breanna~M
  Duffy, Zhaohui Huang, Lauren~D Black, and Brian~P Timko.
\newblock Heart-on-a-chip model with integrated extra-and intracellular
  bioelectronics for monitoring cardiac electrophysiology under acute hypoxia.
\newblock {\em Nano Letters}, 20(4):2585--2593, 2020.

\bibitem{blankenship2023complex}
Brian~W Blankenship, Zachary Jones, Naichen Zhao, Harpreet Singh, Adrisha
  Sarkar, Runxuan Li, Erin Suh, Alan Chen, Costas~P Grigoropoulos, and Ashok
  Ajoy.
\newblock Complex three-dimensional microscale structures for quantum sensing
  applications.
\newblock {\em Nano Letters}, 23(20):9272--9279, 2023.

\bibitem{yan2014micro}
Li~Yan, Jinfeng Zhang, Chun-Sing Lee, and Xianfeng Chen.
\newblock Micro-and nanotechnologies for intracellular delivery.
\newblock {\em Small}, 10(22):4487--4504, 2014.

\end{thebibliography}
 \bibliographystyle{unsrt}
\end{document}